\documentclass[12pt]{article}
\usepackage{amsmath}
\usepackage{amssymb}

\date{}

\title{CLASSICAL SIGNAL MODEL FOR QUANTUM CHANNELS}

\author{Andrei Khrennikov\\nternational Center for Mathematical Modelling
\\in Physics and Cognitive Sciences,\\
Linnaeus University,  V\"axj\"o, S-35195,  Sweden\\
Masanori Ohya and Naboru Watanabe\\Department of Information Sciences, Tokyo University of Science\\
Yamasaki 2641, Noda-shi, Chiba, 278-8510 Japan}

\begin{document}

\maketitle

\begin{abstract}\noindent
Recently it was shown that the main distinguishing features of quantum mechanics (QM) can be reproduced by
a model based on classical random fields, so called  prequantum classical statistical field theory (PCSFT).
This model  provides a possibility to represent averages of quantum observables, including correlations of observables on subsystems of a composite system (e.g., entangled  systems), as averages with respect to fluctuations of classical (Gaussian) random fields.
In this note we consider some consequences of PCSFT for quantum information theory. They are based on the observation \cite{W}
of two authors of this paper  that classical Gaussian channels  (important in classical signal theory) can be represented as
quantum channels. Now we show that quantum channels
can be represented as classical linear transformations of classical Gaussian  signals.
\end{abstract}

\medskip

\section{Introduction}

At the very beginning of QM the idea that quantum mechanics is simply a special model of {\it wave mechanics} was quite popular.
It was supported by the discovery of ``quantum wave equation'' by Schr\"odinger and by the association with each particle its wave-length, the De Broglie wave-length. However,  both Schr\"odinger and De Broglie should give up in front of the difficulties of the interpretation of
``quantum waves'' as real physical waves, see \cite{AKK6} for debates. The main problem of the ``physical wave interpretation'' was the impossibility to describe a composite system by waves defined on the physical space, $X = {\bf
 R}^3.$ The wave function of a composite system is
defined on the space $X_m = {\bf R}^{3m},$ where $m$ is the number of subsystems.
Nevertheless,  nowadays various wave models are  popular in
attempts to go beyond QM. At the present time the most successful models are stochastic
electrodynamics (SED) \cite{R2}, and semiclassical theory \cite{C1}.

Recently \cite{PCSFT1}-- \cite{PCSFT3} it was shown that the main distinguishing features of quantum mechanics (QM) can be reproduced by
a model based on classical random fields, so called  prequantum classical statistical field theory (PCSFT).
This model  provides a possibility to represent averages of quantum observables, including correlations of observables on subsystems of a composite system (e.g., entangled  systems), as averages with respect to fluctuations of classical (Gaussian) random fields.
In this note we consider some consequences of PCSFT for quantum information theory. They are based on the observation \cite{W}
of two authors of this paper  that classical Gaussian channels  (important in classical signal theory) can be represented as
quantum channels. Now we show that quantum channels
can be represented as classical linear transformations of classical Gaussian  signals.

Finally, we mention another mathematical model reducing quantum randomness to classical one, namely, tomographic approach,
see, e.g.,  \cite{M1}. It is an interesting (but may be quite complicated) problem to analyze the interplay between quantum and
classical information theories in this approach.

\section{Brief presentation of PCSFT}

\subsection{Classical fields as hidden variables}

Classical fields are  selected as hidden variables.
Mathematically these are functions $\phi: {\bf R}^3 \to {\bf C}$
(or more generally $\to {\bf C}^k)$ which are square integrable,
i.e., elements of the $L_2$-space.

The latter condition is
standard in the classical signal theory; in particular, for the
electromagnetic field this is just the condition that the energy
is finite
\begin{equation}
\label{SEA5}
{\cal E}(\phi)= \int_{{\bf R}^3} (E^2(x) +B^2(x)) dx < \infty
\end{equation}
or by using is  the Riemann-Silbertstein
vector
$
\phi(x)=  E(x) + iB(x): \Vert \phi \Vert^2 = \int_{{\bf R}^3} \vert
\phi(x) \vert^2 < \infty.
$
Thus the state space of our prequantum model is
$H=L_2({\bf R}^3; {\bf C}),$ the space of square integrable complex valued fields on ``physical space''
${\bf R}^3.$ Formally, the same state space is used in QM, but we
prefer to emphasize coupling with the classical signal theory.
For example, the quantum wave function should satisfy the normalization
condition $\Vert \psi \Vert^2= 1,$
but a PCSFT-state can be any vector of $H.$

\medskip

A random field (at the fixed instant of time)  is a function
$\phi(x, \omega),$  where $\omega$ is the random parameter. Thus,
for each $\omega_0,$ we obtain the classical field, $x\mapsto
\phi(x, \omega_0).$ Another picture of the random field is the
$H$-valued random variable, each $\omega_0$ determines a vector
$\phi(\omega) \in H.$ A random field is given by a probability
distribution on $H.$  To simplify considerations, we  can consider
a finite-dimensional Hilbert space, instead of $L_2({\bf R}^3; {\bf C})$
(as people often do in quantum information theory). In this case
our story is on $H$-valued random vectors, where $H={\bf C}^n.$
This is the ensemble model of the random field. In the rigorous
mathematical framework it is based on {\it Kolmogorov probability
space}  $(\Omega, {\cal F}, {\bf P}),$
where $\Omega$ is a set and ${\cal F}$ is a $\sigma$-algebra of its subsets,
${\bf P}$ is a probability measure on ${\cal F}.$
As  is well known from the classical signal theory, one
can move from the ensemble description of randomness to the time
series description -- under the ergodicity hypothesis.

By applying a linear functional $y$ to the
random vector $\phi$ we obtain the scalar random variable. In the
$L_2$-case we get a family of scalar random variables:
$$
\omega \mapsto \xi_y(\omega) \equiv \int y(x) \overline{\phi(x,\omega)} dx, y \in L_2.
$$
We recall that the covariance operator $D$ of a random field (with zero average)  $\phi \equiv  \phi(x,\omega)$ is defined by its bilinear form:
\begin{equation}
\label{COV}
\langle Du, v\rangle = E \langle u, \phi \rangle  \langle \phi, v \rangle, u, v\in H.
\end{equation}
Under the additional assumption that the prequantum random fields
are {\it Gaussian}, the covariance operator uniquely determines
the field.  We shall poceed under the {\it assumption that prequantum fields
are Gaussian.}
We also suppose that that all {\it prequantum fields have zero average:}
\begin{equation}
\label{SEA6P} E\langle y, \phi \rangle= 0, y \in H,
\end{equation}
where $E$ denotes the classical {\it mathematical expectation}
(average, mean value).
If $H={\bf C}^n,$ where $\phi(\omega)=
(\phi_1(\omega), ...,\phi_n(\omega)),$ then condition
(\ref{SEA6P}), zero average, is reduced to $E\phi_i=0, i=1,2,$ and
the covariance matrix $D= (d_{kl}),$ where $d_{kl}=E\phi_k \bar{\phi}_l.$

\subsection{Covariance operator interpretation of wave function}

In our model the wave function $\psi$ of the QM-formalism
encodes a prequantum random  field: $\phi\equiv \phi_\psi.$ The
QM-terminology, ``a quantum system in the state $\psi$'', is
translated into the PCSFT-terminology, a ``random field.'' In PCSFT
the {\it $\psi$-function determines the covariance operator of the
prequantum random field.} For simplicity, we consider the case of
a single, i.e., noncomposite, system, e.g., photon or electron.
In this situation  normalization (by dispersion) of the covariance operator $D$ of the prequantum field
is given by the orthogonal projector
on the vector $\psi$ (the density operator corresponding to this pure state)
\begin{equation}
\label{SEA6}
\rho_\psi= \psi \otimes \psi,
\end{equation}
i.e., $\rho_\psi u = \langle u, \psi\rangle \psi, \; u \in H.$
The covariance operator of the prequantum field is given by
\begin{equation}
\label{SEA67}
D= \sigma^2(\phi) \rho_\psi,
\end{equation}
where
\begin{equation}
\label{SEA3j2}
 \sigma^2(\phi)= E \Vert \phi \Vert^2 (\omega)= \rm{Tr} D
\end{equation}
is the dispersion of the prequantum random field $\phi \sim N(0,D).$  To determine the covariance operator $D$ on the basis of the density operator $\rho_\psi,$ one should determine the scale of fluctuations of the prequantum field given by the dispersion $\sigma^2(\phi).$

\subsection{Quantum observables from quadratic forms of the prequantum
field}

In PCSFT quantum observables are represented by corresponding
quadratic forms of the prequantum field.  A self-adjoint operator
$\widehat{A}$ is considered as the symbolic representation of the
PCSFT-variable:
\begin{equation}
\label{SEA2}
\phi \mapsto f_A(\phi) = \langle \widehat{A} \phi, \phi \rangle.
\end{equation}
We remark that $f_A$ can be considered as a function on the phase space
of classical fields: $f_A\equiv f_A(q,p),$ where $\phi(x)=q(x)+ip(x),
q, p \in L_2({\bf R}^3; {\bf R}),$ the space of real valued fields.

Consider the quantum average
\begin{equation}
\label{SEA1}
\langle f_A \rangle_{{\rm{QM}}} =\langle \widehat{A} \psi, \psi \rangle.
\end{equation}
and the classical average
\begin{equation}
\label{SEA1k}
\langle f_A \rangle_{{\rm{CL}}}= Ef_A(\phi)
\end{equation}
They coincide up to a scaling factor
\begin{equation}
\label{SEA1l}
\langle f_A \rangle_{{\rm{CL}}}= \rm{Tr} D \; \langle f_A \rangle_{{\rm{QM}}}.
\end{equation}

In the
real physical case $H$ is infinite-dimensional;  the classical average is given by the integral
over all possible classical fields; probabilistic weights of the fields are
determined by the $\psi.$

\subsection{Quantum and prequantum interpretations of
Schr\"odinger's equation} \label{SER}

{\bf Main message:} {\it Schr\"odinger's equation does not describe the dynamics of a wave of probability.
The same equation plays a double role. On the one hand, it describes dynamics of a physical random field.
On the other hand, it encodes the dynamics of the covariance operator of this field.}

\medskip

Before to go to the PCSFT-dynamics, we consider the Schr\"odinger equation in the standard
QM-formalism:
\begin{equation}
\label{SE}
i h \frac{\partial \psi}{\partial t}(t, x)= \widehat{\cal H} \psi (t, x),
\end{equation}
\begin{equation}
\label{SE1}
\psi(t_0, x)=\psi_0(x),
\end{equation}
where $\widehat{\cal H}$ is Hamiltonian, the energy observable. Although
Shcr\"odinger tried to interpret $\psi(t,x)$ as  a classical field
(e.g., the electron field), the conventional interpretation
is the probabilistic one, due to Max Born.

\medskip

We recall that a time dependent random field $\phi(t, x, \omega)$  is called the {\it stochastic
process}  (with the state space $H)$. The dynamics of the prequantum random field
 is described by the simplest stochastic process which is given by {\it deterministic dynamics with
 random initial conditions.}

In PCSFT the Schr\"odinger equation, but with the random initial condition, describes
the dynamics of the prequantum random field, i.e., the prequantum stochastic process can be obtained
from the same mathematical equation as it was used in QM for the dynamics of the wave function:
\begin{equation}
\label{SE2}
i h \frac{\partial \phi}{\partial t}(t, x, \omega)= \widehat{\cal H} \phi (t, x, \omega),
\end{equation}
\begin{equation}
\label{SE3}
\phi(t_0, x, \omega)=\phi_0(x, \omega),
\end{equation}
where the initial random field $\phi_0(x, \omega)$ is determined
by the quantum pure state $\psi_0 \sim N(0,D),$ where $D$ is given by (\ref{SEA67}).
 Hence, the standard QM provides the knowledge
of the covariance operator of the prequantum random field.

The PCSFT dynamics (\ref{SE2}), (\ref{SE3}) matches with the
standard QM-dynamics (\ref{SE}), (\ref{SE1}) -- by taking into
account the PCSFT-interpretation of  the wave-function, see
(\ref{SEA6}). Set  $$\rho(t)= D(t)/\rm{Tr}D(t),$$ where $D(t)$ is  the covariance operator of the
random field $\phi(t)\equiv \phi(t,x,\omega),$  the solution of
(\ref{SE2}), (\ref{SE3}). Then $$\rho(t)\equiv \rho_{\psi}(t)=
\psi(t)\otimes \psi(t),$$ where $\psi(t)$ is a solution of
(\ref{SE}), (\ref{SE1}).

\subsection{Random fields corresponding to mixed states}

We now consider the general quantum state
given by a density operator $\rho.$
By PCSFT it is determined  as normalization of the
covariance operator of  the corresponding prequantum field
\begin{equation}
\label{L1}
 \rho = D/\rm{Tr}\; D.
\end{equation}
The dynamics  of  the prequantum field $\phi(t,x,\omega)$ is also
described by the Shr\"odinger equation, see (\ref{SE2}), (\ref{SE3}),  with the random initial
condition $\phi_0(x,\omega) \sim N(0,D).$

\section{Linear transformations of Gaussian random fields}

Let $H_i, i=1,2,$ be complex Hilbert spaces. Consider a linear
bounded operator $V: H_1 \to H_2$ (we remark that $V^*: H_2 \to
H_1)$ and the corresponding linear filter
\begin{equation}
\label{LF} \phi_{\rm{out}}(\omega)= V \phi_{\rm{in}}(\omega),
\end{equation}
where $\phi_{\rm{in}}$ is the $H_1$-valued random field which is
ditributed $N(0,D_{\rm{in}}).$ Then the $H_2$-valued random field
$\phi_{\rm{out}} \sim N(0, D_{\rm{out}}),$ where
\begin{equation}
\label{LF1} D_{\rm{out}}= V D_{\rm{in}} V^*.
\end{equation}

\medskip

{\bf Example 1.} (Unitary evolution) The solution of the
Schr\"odinger equation (\ref{SE2}) with the initial condition
$\phi_{\rm{in}},$ a Gaussian random field, can be represented in
the form of the linear filter (\ref{LF}), where $V=U_t=\exp\{-it
\widehat{{\cal H}}/h\}.$ This filter preserves the norm of the
prequantum field
\begin{equation}
\label{LF3} \Vert \phi_{\rm{out}}(t, \omega) \Vert^2= \Vert
\phi_{\rm{in}}(\omega) \Vert^2.
\end{equation}
Hence, this filter preserves even the dispersion of a classical
signal
\begin{equation}
\label{LF4}
 \sigma^2(\phi_{\rm{out}})(t) = \rm{Tr} D_{\rm{out}}(t) =   \sigma^2(\phi_{\rm{in}})=  \rm{Tr} D_{\rm{in}}.
\end{equation}

\medskip

{\bf Example 2.} (von Neumann-L\"uders projection). Let $L=P$ be
an orthogonal projector $P:H \to E,$ where $E$ is a linear
subspace of $H.$  Here the $E$-valued (Gaussian) random field $
\phi_{\rm{out}}(\omega) = P \phi_{\rm{in}}(\omega) $ describes the
output after the von Neumann-L\"uders projection-measurement (with
filtration with respect to the value corresponding to the
projector $P).$  The the von Neumann-L\"uders filter does not
preserve even the dispersion $$\rm{Tr} D_{\rm{out}}=\rm{Tr} P
D_{\rm{in}}P \not= \rm{Tr} D_{\rm{in}}.$$  Moreover, $$\rm{Tr}
D_{\rm{out}} \leq \rm{Tr} D_{\rm{in}},$$ i.e., the dispersion of a
signal always decreases.

\medskip

{\bf Example 3.} (von Neumann-L\"uders measurement). Consider a
dichotomous quantum observable $\widehat{A}$ with eigenvalues
$\alpha_1$ and $\alpha_2$ and eigensubspaces $E_1$ and $E_2.$
Denote corresponding projectors by $P_1$ and $P_2.$ They are
orthogonal. Take two {\it independent random fields} $\phi_{k}
\sim N(0,D), k=1,2,$ where the corresponding quantum state $\rho=
D/\rm{Tr} D.$  We form the $H\times H$ valued random variable
$\phi_{\rm{in}}(\omega) =(\phi_{1}(\omega), \phi_{2}(\omega)) \sim
N(0, D\times D).$ We define a linear operator $V: H \times H \to
H, V(x,y)= P_1 x + P_2 y$ and the corresponding linear filter
\begin{equation}
\label{LF6} \phi_{\rm{out}}(\omega)= V \phi_{\rm{in}}(\omega) =
P_1 \phi_{1}(\omega) + P_2\phi_{2}(\omega).
\end{equation}
Since $\phi_{1}$ and $\phi_{2}$ are independent and Gaussian,
their linear transforms $P_1 \phi_{1}$ and $ P_2 \phi_{2}$ are
also independent. Hence, the covariance operator of the random
field $\phi_{\rm{out}}$ equals to the sum of covariance operators
of the latter two random fields, i.e.,
\begin{equation}
\label{LF7} D_{\rm{out}}= P_1 D_{\rm{in}} P_1 + P_2 D_{\rm{in}}
P_2.
\end{equation}
We remark that the dispersion is preserved
\begin{equation}
\label{LF4}
 \rm{Tr} D_{\rm{out}}=  \rm{Tr} D_{\rm{in}}.
\end{equation}

This example can be easily generalized to an arbitrary quantum
observables with purely discrete spectrum, $\widehat{A}= \sum_k
\alpha_k P_k;$ moreover, to any POVM. For simplicity, we present
the case of  measurement with finitely many results.

\medskip

{\bf Example 3.} (POVM-measurement as classical linear filter)
Consider a POVM $\{Q_i=V_i V_i^*\}_{i=1}^n,$ where
\begin{equation}
\label{LF8} \sum_{i=1}^n Q_i=I.
\end{equation}
Define the corresponding linear map $V: H_1 \times ... \times H_n
\to H, V (x_1,...,x_n) = V_1 x_1 +...+ V_n x_n.$ Consider also  a
quantum state, density operator $\rho.$ Take a vector
$\phi_{\rm{in}}(\omega) =(\phi_{1}(\omega),..., \phi_{n}(\omega))$
consisting of equaly distributed independent Gaussian random
fields of $N(0,D)$-type, where the quantum state under
consideration $\rho= D/\rm{Tr} D.$ We define a linear filter
corresponding to the map $V$
\begin{equation}
\label{LF9} \phi_{\rm{out}}(\omega)= V \phi_{\rm{in}}(\omega) =
\sum_i V_i \phi_{i}(\omega).
\end{equation}
We have
\begin{equation}
\label{LF10} D_{\rm{out}}= \sum_i V_i D_{\rm{in}} V_i^*.
\end{equation}
This is nothing else than a completely positive map, see, e.g.,
\cite{O}. Hence, it preserves the trace, i.e., the equality
(\ref{LF4}) holds. Therefore the equality (\ref{LF10}) for
covariance operators is transformed into the equality for quantum
states, density operators
\begin{equation}
\label{LF12} \rho_{\rm{out}}=  \frac{D_{\rm{out}}}{\rm{Tr}
D_{\rm{out}}} =  \frac{D_{\rm{out}}}{\rm{Tr} D_{\rm{in}}}=  \sum_i
\frac{\rm{Tr} V_i D_{\rm{in}} V_i^*}{\rm{Tr} D_{\rm{in}}}
 \rho_{\rm{out},i},
\end{equation}
where
\begin{equation}
\label{LF13}
 \rho_{\rm{out},i} = \frac{V_i D_{\rm{in}} V_i^*}{\rm{Tr} V_i D_{\rm{in}} V_i^*}.
\end{equation}
Set $\rho_{\rm{in}}= D_{\rm{in}}/\rm{Tr}D_{\rm{in}}.$ Then
\begin{equation}
\label{LF14}
 \rho_{\rm{out},i} = \frac{V_i \rho_{\rm{in}} V_i^*}{\rm{Tr} V_i \rho_{\rm{in}} V_i^*}
\end{equation}
and
\begin{equation}
\label{LF15} \rho_{\rm{out}}=\sum_i (\rm{Tr} V_i \rho_{\rm{in}}
V_i^*) \;
 \rho_{\rm{out},i}.
\end{equation}

Since any {\it quantum channel} can represented (the Kraus
representation,  \cite{O}) in the form (\ref{LF10}), in general
with the infinite number of terms, we demonstrated (generalization
to the infinite number of terms in the Kraus decomposition is
straightforward) that any quantum channel can be represented as
the linear filter of classical signals of the form (\ref{LF9}).
This is a step in the same direction as paper \cite{W}: exploring
the analogy between quantum information theory and classical
signal theory.

This research project on the interplay between classical and quantum information theory
started during the visiting fellowship of A. Khrennikov at Tokyo University of Science,
March 2010, which was supported by the grant QBIC.

\end{document}